# Site-specific stable deterministic single photon emitters with low Huang-Rhys value in layered hexagonal boron nitride at room temperature

Amit Bhunia[1*], Pragya Joshi[2*], Nitesh Singh[1], Biswanath Chakraborty[2]
and Rajesh V Nair[1]

[1]*Laboratory for Nano-scale Optics and Meta-materials (LaNOM), Department of Physics, Indian Institute of Technology Ropar 140001, India*

[2]*Department of Physics, Indian Institute of Technology Jammu, Jagti, NH-44, Jammu and Kashmir 181221, India*

**ABSTRACT:**

Development of stable room-temperature bright single-photon emitters using atomic defects in hexagonal-boron nitride flakes (h-BN) provides significant promises for quantum technologies. However, an outstanding challenge in h-BN is creating site-specific, stable, high emission rate single photon emitters with very low Huang-Rhys (HR) factor. Here, we discuss the photonic properties of site specific, isolated, stable quantum emitter that emit single photons with a high emission rate and unprecedented low HR value of 0.6±0.2 at room temperature. Scanning confocal image confirms site-specific single photon emitter with prominent zero-phonon line at ~578 nm with saturation photon counts of $3.5 \times 10^5$ counts/second. The second-order intensity-intensity correlation measurement shows an anti-bunching dip of ~0.25 with an emission lifetime of 2.46±0.1 ns. The importance of low-energy electron beam irradiation and subsequent annealing is emphasized to achieve stable single photon emitters.

*contributed equally

# I. INTRODUCTION:

The stable, deterministic, room-temperature, bright single photon sources are the cornerstone for photonic quantum technologies, and subsequent manipulation of their quantum states is required for quantum information processing, quantum sensing, and spin-photon interface. In this context, many single photon sources like cold atoms,[1] quantum dots (QDs),[2] color centers in solids, and non-linear parametric down-conversion have been tested in recent times.[3] Color centers in diamond and silicon carbide are promising single-photon sources at room temperature.[4] The emission spectra of nitrogen-vacancy (NV) centers in diamonds consist of pure electronic transitions at zero phonon line (ZPL) and significant phonon sideband (PSB) emission due to phonons with limited emission into ZPL. Such broad PSB emission can induce decoherence, limiting diamond use in quantum applications. Hence, PSB suppression is required with simultaneous ZPL enhancement, which is quantified using Debye-Waller (DW) and Huang-Rhys (HR) factors. An enhanced DW factor and low HR factor are advantageous to obtain indistinguishable single photon sources at ZPL so that the entanglement rate and efficient spin-photon interface required for the long-distance quantum network can be achieved. Hence, easy-to-fabricate, room-temperature, bright single photon emitters with distinct ZPL with low HR factors are envisioned. The two-dimensional (2d) van der Waals materials including transition-metal dichalcogenides (TMDs) have drawn significant attention as single photon sources.[5] Although TMDs have demonstrated evidence of single photon properties, their inherent low bandgap and low-temperature requirement[6] limit their suitability in quantum applications. Hexagonal boron nitride (h-BN) is a viable 2d layered system with a bandgap of 6 eV that can overcome the shortcomings of TMDs-and diamond-based single photon emitters. The h-BN consists of a hexagonal arrangement of Boron and Nitrogen atoms in a 2d plane. The absence of B or N atoms, the interchange of their positions, and the intrusion of foreign atoms can act like sub-band gap point defects.[7-9] Excellent thermal and chemical stability of h-BN enables its use in any robust environment and exhibits unique properties of natural hyperbolic dispersion, enhancing light-matter interaction.[10] It is easy to integrate h-BN layers with photonic cavities/waveguides so that extraction efficiency and directionality of emitted single photons can be significantly improved.[11,12] Also, h-BN can host a variety of point defects with respective emission wavelengths spanning from visible to near-infrared range.[13,14] However, only a few of them are experimentally studied like $V_NN_B$, $V_NC_B$, $V_BC_N$, $V_B^-$ and a detailed understanding of their structure and spin properties remain elusive.[15,16] Usually, the defects are inherently present or incorporated during h-BN sample growth. The carbon presence is ubiquitous as an omnipresent foreign atom that naturally appears in h-BN. Hence, such graphite signature is observed in the Raman spectra of h-BN sample.[9,17-19] The graphite Raman peaks are often identified with ZPL; therefore, care should be taken while interpreting emission peaks. Additionally, the defect complexes can be created by irradiating h-BN flakes with low-energy electrons, energetic pulsed lasers, and annealing in an inert atmosphere to stabilize defects.[20-,22] Recently, spin properties of $V_B^-$ defects in h-BN have been studied using optically detected magnetic resonance[23-25]. Strain in h-BN can play a substantial role in tuning the emission properties of quantum emitters[18,26]. Eventually, a few layers of h-BN must be integrated with photonic structures for precise quantum states manipulation to achieve an effective spin-photon interface.[27] However, the challenge remains to achieve easily traceable site-specific 'single' stable single photon emitters in h-BN without any crystal deformation and cumbersome fabrication techniques. The other challenge is eliminating rapid fluctuations in spectral lines, known as spectral diffusion,[28] with stable photon counts and low HR values. For next-generation advanced quantum technologies, such single-photon emitter fidelity in samples with very low HR factor is an essential requirement.

Here, we report site-specific, 'single' photo-stable, deterministic single photon emitters in different h-BN flake that show excellent single photon purity with low HR factor at room temperature. The importance of low-energy irradiation and subsequent annealing to stabilize single photon emitters is elaborated. The measured emission spectra confirm a prominent ZPL peak distinct from PSB emission. The DW factor is estimated to be 50±10% and the HR factor 0.6±0.2 which is much less than unity, indicating minimal phonon involvement in the emission process. The photon anti-bunching measurements show a dip at zero-time delay with a value of 0.25, indicating a high-purity single photon with an emission lifetime of 2.46 ±0.1 ns, corroborated by independent lifetime measurements.

## II. RESULTS AND DISCUSSIONS

### IIA. Sample details

Few layers of h-BN were mechanically exfoliated from an h-BN crystal (M/s. 2D Semiconductors, USA) using the standard scotch tape method onto a SiO$_2$/Si substrate with a 285 nm oxide layer (Supporting Information S1). The flakes are irradiated using an electron beam of acceleration voltage 20 kV and probe current of 6 µA in a field emission scanning electron microscope (JEOL) for 20 seconds. The samples are then annealed in argon atmosphere at 850°C for two hours. Sample irradiation for more than 20 seconds could result in graphitization of the sample. Figure 1a shows the schematic of the h-BN monolayer consisting of Boron and Nitrogen atoms arranged in a hexagonal symmetry pattern. The formation of a complex defect state due to probable presence of carbon atoms that replace Boron or Nitrogen atoms is shown. Many such carbon-related defect complexes exist in h-BN[13]; only a few show reproducible single photon emission. Our experimental results agree to some extent with the estimated ZPL energy corresponding to $C_2C_N$ defect complex (Figure 1a) and consequently, we will discuss the role of this defect complex in generating single photons.[29] We consider three different types of h-BN samples in the present work: (i) 'Irradiated and annealed' flakes (*F1* sample), i.e. the sample is irradiated using an electron beam to generate defects and subsequently annealed at high temperature to stabilize the defects, (ii) 'only annealed' flakes (*F2* sample), which is annealed only and not irradiated, and (iii) 'pristine sample, which is not irradiated or annealed' flakes (*F3* sample) [Supplementary Information: S1]. Figure 1b shows the optical image of *F1* sample using an objective of magnification 50X (Nikon Eclipse). Figures 1c and 1d represent surface morphology and height profile of *F1* sample (flakes seen in Figure 1b) using an atomic force microscope operating in the tapping mode (MultiMode 8, Bruker). The height profile measured at two extreme locations on the sample is given in Figure 1d, which shows a thickness variation from 25 nm to 30 nm (green and blue dashed lines) corresponding to 75 to 90 layers. The thickness of layered h-BN has a crucial impact on the characteristics of single photon emission properties.[7] However, we focus on h-BN samples having identical average thickness and therefore, expect similar emission properties.

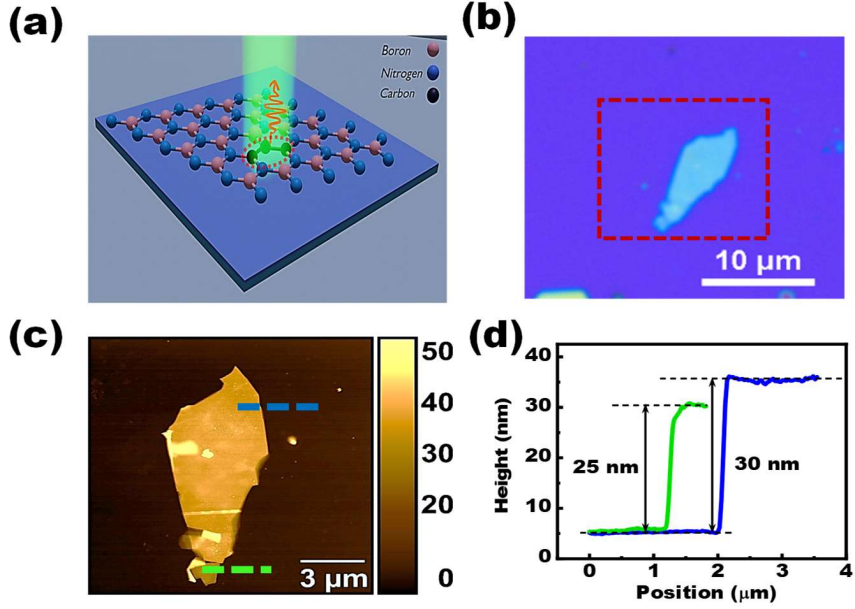

Figure 1. (a) Schematic representation of a typical h-BN structure consists of Boron and Nitrogen atoms in a hexagonal honeycomb pattern. A carbon-related defect complex $C_2C_N$ is depicted for better realization (b) The optical microscope image of the *F1* sample on Si/SiO$_2$ substrate is captured using a 50X microscope objective. (c) The AFM image of *F1* sample is shown in (b). The sample size is approximately $12 \times 3$ μm$^2$ inferred from the AFM image. (d) The height profiles obtained from AFM images at two locations show thickness values of 30 nm (dashed blue line) and 25 nm (dashed green lines), respectively.

**IIB. Confocal imaging and emission measurements**

To assess single photon emitters, it is required to identify spatial locations of emitter through scanning confocal microscopy (Supporting Information: S2). Figure 2a represents a room-temperature scanning confocal microscopy image of *F1* sample using a 532 nm continuous wave (CW) laser. We identify a bright red spot with a dimension of ~1μm$^2$ in the confocal image with photon counts much higher than the surroundings, confirming an isolated quantum emitter ($Q_1$ point) might be present on the *F1* sample. The $Q_1$ point is identified near the flake edges.[22,30] Figure 2b shows the measured emission spectra from $Q_1$ point which shows a prominent peak at $578 \pm 2$ nm along with a well-separated, seemingly broad spectral hump centered at $626 \pm 5$ nm. The intense peak at 578 nm is identified as the ZPL transition, whereas the ~50 nm (~165 meV) red-shifted broad peak at 626 nm is identified as PSB. The ZPL-related peak is fitted with the Lorentzian function and the estimated full width at half maxima is $15 \pm 0.2$ nm (Supplementary Information: Figure S9). The presence of a few layers of h-BN is confirmed by characteristic Raman mode at 1365 cm$^-$

[1] with Raman shift energy of ~166 meV using a 532 nm excitation laser (Supporting Information: S3). The estimated Raman shifted energy agrees with the energy difference corresponding to PSB and ZPL.[31] Thus, ZPL transitions and accompanying PSB emission are well connected with Raman-related vibronic level transition, implying direct involvement of optical phonons. To substantiate the distinct photonic properties of *F1* sample, we have thoroughly examined multiple samples in the *F1* category (Supporting Information: S5). We have measured similar photonic properties from the stable, isolated, and deterministic quantum emitters in multiple h-BN sample. Further, by focussing the electron beam at different spatial location on the sample, site-specific stable single photon emitters can be created in h-BN samples. The measured ZPL wavelength is intact at ~576±2 nm for all the samples, which indicates the same kind of defect complexes are created in all the sample in the *F1* category. Moreover, all these quantum emitters show a long term spectral stability as evidenced by the nominal peak shift of ±2 nm (Supporting Information: Table S1).

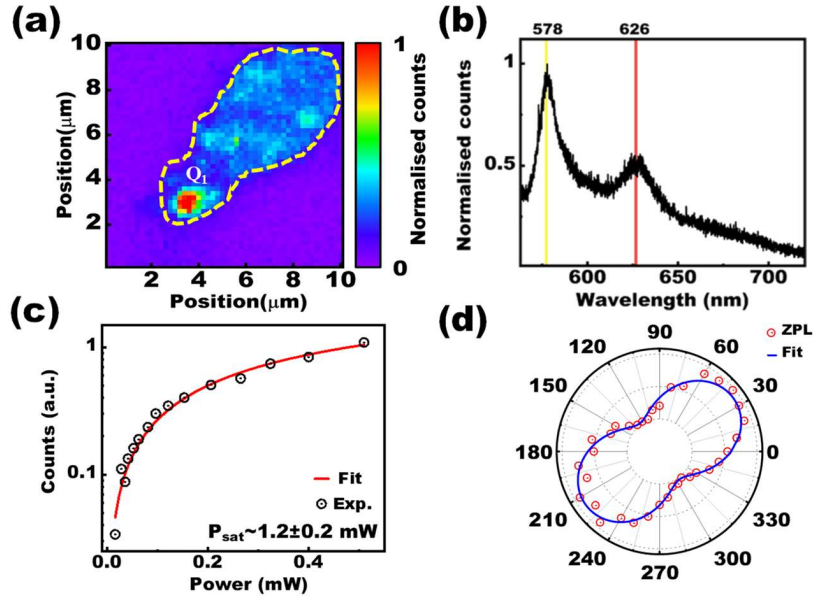

**Figure 2**. (a) The scanning confocal image of the *F1* sample (dotted line) hosts an isolated single photon emitter (red spot). It is seen that only a site-specific single bright spot is observed throughout the flake ($Q_1$ point) (b) Normalized emission spectra depict prominent ZPL (yellow line) along with broader PSB emission (red line) red-shifted by ~ 165 meV. (c) The integrated photons count (symbols) corresponds to the ZPL emission peak as a function of the excitation power of the laser. It shows the saturation behaviour for the quantum emitter $Q_1$ at high excitation power. (d) Linear polarization-dependent dipolar properties of quantum emitter measured at ZPL wavelength. The measured emission intensity (symbols) is fitted with $cos^2\theta$ function (solid line), which gives the polarization angle of the emitter as $36\pm1^0$. The degree of polarisation is estimated to be ~ 47% for the ZPL-related emission.

Figure 2c depicts the integrated emission intensity ($I$) corresponding to ZPL peak (in a semi-logarithmic scale), which shows saturation with an increase in excitation power ($P$). The experimental results follows the power-dependent phenomenological equation: $I = I_\infty \frac{P}{P+P_0}$, where $I_\infty$ and $P_0$ are the emission intensity counts and excitation power at saturation. The estimated value of $P_0$ from fit is 1.2±0.2 mW with a saturation photon count of 3.5±0.6 ×$10^5$ counts/second. We observe a prominent signature of dipolar emission for ZPL emission peak as shown in Figure 2d, implying an isolated single quantum emitter at $Q_1$ point. The excitation laser polarization is fixed along the linear polarizer optic axis, and analyzer optic axis is rotated to measure emission intensity. Integrated emission intensity as a function of analyser angle is plotted while the excitation laser polarization is kept fixed at the optic axis of the polarizer. The polarization-dependent angular emission intensity follows $cos^2\theta$ dependence, where '$\theta$' is the angle between emitted light polarization and analyzer optic axis. The measured angular emission intensity variation is fitted using: $I(\theta) = A + Bcos^2\left(\frac{\pi}{180}(\theta - \langle\theta\rangle)\right)$ to estimate dipole orientation.[32] Here $\langle\theta\rangle$ represents average polarization angle of emitted dipoles and $I(\theta)$ is the integrated emission intensity of emitter. The measured value of $\langle\theta\rangle$ for ZPL emission peak is ~36±$1^0$. Furthermore, to understand the degree of polarization of the emitter, the visibility is calculated as $\left(\frac{I_{max} - I_{min}}{I_{max} + I_{min}}\right)$, where $I_{max}$ and $I_{min}$ represent maximum and minimum integrated emission intensity value, obtained from Figure 2d. The estimated higher visibility value of ~47% indicates the presence of distinct and isolated dipoles in the system. However, in contrast to the known dipolar behavior of the quantum emitter, the PSB-related emission peak shows better dipolar behavior and a higher degree of polarization than ZPL emission peaks. In addition, it is possible to obtain pure ZPL emission by suppressing the PSB-related emission by precise adjustment of analyzer angle (Supporting Information: S4).

## IIC. Single photon emission studies

To confirm single photon emitters in *F1* sample, we measure the second-order intensity-intensity correlation function ($g^{(2)}(\tau)$) using a home-built Hanbury Brown-Twiss set-up (Supporting Information: S1). The $g^{(2)}(\tau)$ curve is measured using 532 nm CW laser with 0.5 *mW* excitation power at the ZPL wavelength.

We observe a prominent dip at the zero-time delay ($\tau=0$) with a value of 0.25 in the background corrected $g^{(2)}(\tau)$ curve from $Q_1$ point as seen in Figure 3a (upper panel). Such a substantial dip with $g^{(2)}(\tau) \ll 0.5$ provides an unambiguous single photon signature at room temperature. The $g^{(2)}(\tau)$ curve width is relatively narrow compared to the $g^{(2)}(\tau)$ curve width from single NV centers in diamond; hence, we anticipate bright single photon emission from h-BN flake. The measured $g^2(\tau)$ curve is fitted with exponential function to estimate the emitter lifetime: $g^{(2)}(\tau) = 1 - b\, exp^{-|\tau|/\tau_1}$ with $\tau_1$ as emitter decay rate, $b$ is a fitting parameter, and estimated lifetime is ~ 2.5±0.5 ns. The $g^{(2)}(\tau)$ curve fits well with a single exponential function that implies a dominant two-levels transition process. We emphasize that the decay dynamics from $Q_1$ point is a two-level process which is also seen in the computational results.[29] The $g^{(2)}(\tau)$ curve measurement is repeated using a 532 nm pulsed laser of repetition rate 40 MHz to support generic feature of single photon emitters in *F1* sample. The measured pulsed $g^{(2)}(\tau)$ curve also shows the absence of peak $\tau = 0$ indicating prominent single photon emission, as shown in Figure 3a (lower panel). Further, we have measured the $g^{(2)}(\tau)$ curve from site-specific defects created in multiple samples which belongs to the *F1* category. We find substantial anti-correlation dip below 0.5 implying single photon emission from site-specific emitters in multiple samples (Supplementary Information: S5). The results ascertain site-specific single photon emitters in h-BN at room temperature.

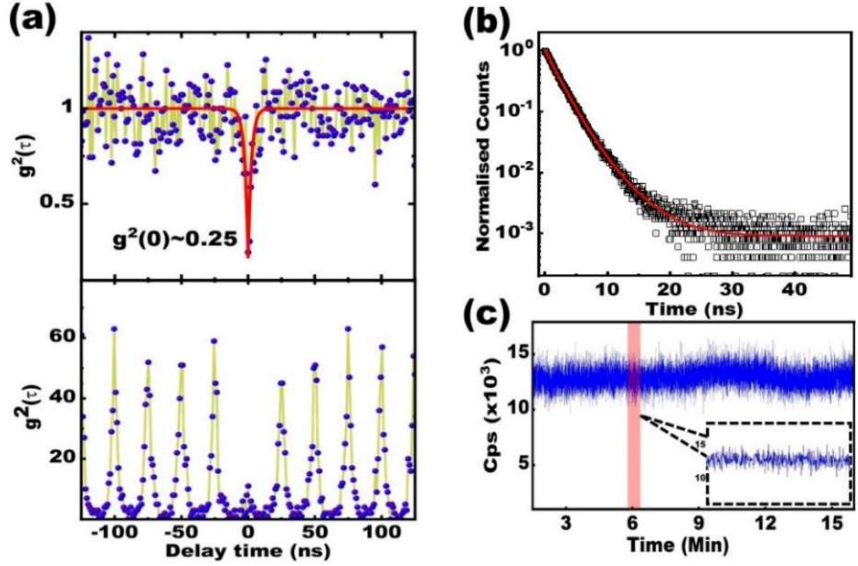

**Figure 3.** (a) The background corrected second-order intensity auto-correlation $g^{(2)}(\tau)$ measurement performed using CW laser (upper panel) shows the photon anti-bunching dip at τ=0 with a value of 0.25, implying a single photon emission. The lower panel shows the photon anti-bunching $g^{(2)}(\tau)$ measurement is repeated using a pulsed laser of repetition rate 40 MHz. The absence of a peak at τ = 0 implies a single photon emission. Both measurements are carried out using a 532 nm laser with a bin size of 1024 ps in the time-axis. (b) The measured excited state decay profile of the quantum emitter using time-resolved emission spectroscopy at the ZPL peak wavelength. The decay curve is fitted (solid red line) using the double exponential function to estimate the average lifetime ($\tau_{avg}$) of 2.46±0.1 ns. (c) The emission counts per second (cps) remain at a fixed value (~12k) for a longer time indicating photon emission stability.

Further to corroborate the fast decay rate of single photon emitters, we independently measured the decay rate using time-resolved emission experiments with a 532 nm pulsed laser with a repetition rate 10 MHz as the excitation source. Figure 3b depicts the measured decay curve (symbol) from $Q_1$ point on *F1* sample fitted with a double exponential function (solid red line) to estimate the average lifetime ($\tau_{avg}$). The $\tau_{avg}$ is calculated using the following expression: $\tau_{avg} = \frac{A_1\tau_1}{A_1\tau_1+A_2\tau_2}\tau_1 + \frac{A_2\tau_2}{A_1\tau_1+A_2\tau_2}\tau_2$, where $A_1$ and $A_2$ are the amplitudes corresponding to lifetime components $\tau_1$ and $\tau_2$, respectively.[33] The estimated $\tau_{avg}$ value for the $Q_1$ point is ~ 2.46±0.1 ns at ZPL, which agrees with $g^{(2)}(\tau)$ measurements, suggesting bright single photon emission. The measured lifetime values from quantum emitters on different samples in the *F1* category shows similar lifetime values, which further support the creation of same kinds of defect complexes in all the samples (Supplementary Information: S5). However, we observe substantially different emission decay curves from other spatial locations compared to $Q_1$ point on the *F1* sample (Supporting Information: S6). The contrasting emission decay curve indicates different decay mechanisms from different spatial locations on *F1* sample such that $Q_1$ point is a unique site-specific single photon emitter. Further, the $Q_1$ point on *F1* sample shows a long-term photo-stability in emission counts for a substantial amount of time, as shown in Figure 3c. The emission counts acquired for fifteen minutes with photon counts per second (cps) remain the same at ~$12 \times 10^3$ counts/second. The stability could be achieved for hours by controlling the experimental conditions. Figure 3c inset shows zoomed view of photon counts (marked with a bar), which demonstrates the absence of photo-blinking or photon-bleaching and stable photon counts even at a tiny time scale.

**IID. Estimation of Debye-Waller and Huang-Rhys factors**

Furthermore, electron-phonon interactions play an essential role in the emission spectra depicted in Figure 2b. Here, we analyse detailed spectral behavior of quantum emitter at $Q_1$ point on the *F1* sample to unravel the interaction between electrons and phonons in terms of DW and HR factors. The ZPL peak is fitted with Lorentzian function indicating homogenous emission process and full-width at half-maximum of ZPL peak is ~15±0.2 nm nm, whereas PSB spectral width from the fit is ~28±0.4 nm. The ZPL-related pure electronic transitions dominate in emission process rather than phonon-mediated PSB transition, which would be useful for quantum technological applications.[16,18] To quantify spectral weightage of ZPL peak in total emission spectra, the DW factor ($\alpha$) is estimated, which is defined as the ratio between integrated emission intensity at ZPL to that of total emission intensity. The estimated $\alpha$ value for $Q_1$ point is 50±10%, much higher than the $\alpha$ value of popularly known NV centers (~3%) at room temperature.[5,34] To further understand electron-phonon interaction, HR factor ($S$) is estimated to provide more intricate details of emitted phonons during electronic transition.[32,35,36] We calculate the $S$ factor from measured $\alpha$ value using relation $S = -\ln(\alpha)$ which results in $S = 0.60 \pm 0.20$. Similar $S$ values are obtained from quantum emitters in other samples of the *F1* category (Supplementary Information: S5). We also estimate the $S$ value following the works by Jungwirth et al.[31]: $\Delta E = E_{exc} - E_{ZPL} = 2S\hbar\omega_0$, where $E_{exc}$ and $E_{ZPL}$ are the excitation laser and ZPL peak energies, respectively, $\hbar$ is Planck's constant, and $\omega_0$ is optical phonon frequency. Considering h-BN optical phonon energy is 170 meV[8,37,38] and $E_{exc}$ = 2.33 eV, the estimated $S$ value is 0.54±0.02. The estimated $S$ value for $F1$ sample by both methods is much lower than many previous reports.[31,39,40] The low $S$ value implies minimal participation of optical phonons during an electronic transition minimizing decoherence. Therefore, emission measurement on $F1$ sample confirms a much purer quantum emitter signature in terms of its single photon emission property.

An obvious question arises here is the type of defect complexes responsible for the single photon emission from the $Q_1$ point in the $F1$ sample. We assign the $Q_1$ defect center with carbon trimer substitutional defect complex $C_2C_N$. We find compelling experimental evidence that relates to the $C_2C_N$ defect complex similar to the computational results.[29] The measured ZPL peak energy 2.14 eV (578 nm) and lifetime (2.46 ns) closely matches with the calculated ZPL and lifetime values for the $C_2C_N$ defect complex. This carbon defects enabled the single photon emission with optical transition $^2_2A_2$-$^2_0A_2$, corroborated by a single exponential fit to the measured $g^{(2)}(\tau)$ curve in Figure 3a (upper panel). The very low estimated HR factor resembles the lowest HR factor values obtained for $C_2C_N$ defect complex in h-BN.[29] The measured Raman peak at 1600 cm$^{-1}$ further confirms the presence of carbon defects in the sample (Figure S2). Our results also confirm the necessity of carbon inclusion in generating stable single photon emitters in h-BN.[9]

**IIE. Emission studies on only-annealed sample ($F2$ sample)**

The quantum nature of light emission from the $F2$ sample is thoroughly explored and subsequently compared with the $F1$ sample. The $F2$ sample thickness is 25±1nm measured using AFM images (Supporting Information: S7). Figure 4a represents the scanning confocal image of the $F2$ sample measured using a 532 nm CW laser. The inset shows optical microscope image confirming the exact sample shape imaged in a confocal microscope.

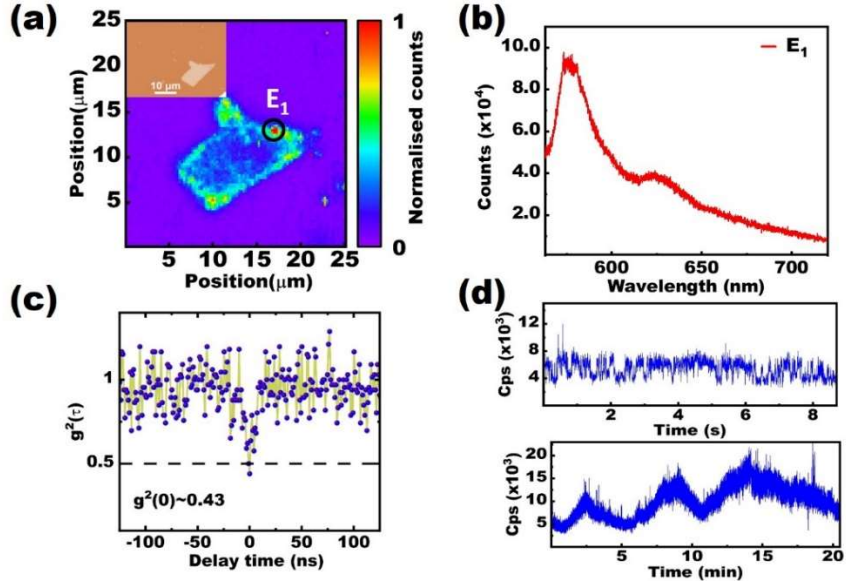

**Figure 4.** (a) The scanning confocal image of the *F2* sample. A bright quantum emitter (encircled with a solid line) near the flake edge is seen as similar to the *F1* sample. The inset depicts the optical image of the same flake taken using a 50X microscope objective. (b) The emission spectra from the $E_1$ point show prominent ZPL at 577 nm and well-separated PSB-related emission. (c) Second-order intensity auto-correlation $g^{(2)}(\tau)$ measurement from $E_1$ point shows photon anti-bunching with a value of 0.43 at $\tau =0$ implying single photon emission. (d) The measured photon counts from the $E_1$ point show a photo-blinking effect as shown in the upper panel. Once the counts are acquired for a longer duration, the measured photon counts show instability, as seen in the lower panel. The $E_1$ point on the *F2* sample is photo-bleached once the excitation source power is increased above a threshold value.

Figure 4b represents emission spectra corresponding to $E_1$ point with ZPL at 577 nm along with well-separated PSB emission centered at 625 nm, similar to the $Q_1$ point on the *F1* sample. We observe a bright spot ($E_1$ point) with counts much higher than the background emission counts. The estimated FWHM of ZPL peak is 22± 0.1 nm while that for PSB emission peak is 50±0.7 nm and both peaks are separated by ~48 nm (165 meV). We do not observe a similarly isolated emitter like $E_1$ point in other regions of *F2* sample, corroborating our previous observation on site-specific, isolated quantum emitter in *F1* sample. Figure 4c shows measured $g^{(2)}(\tau)$ curve from $E_1$ point, which shows a photon anti-bunching dip at $\tau=0$ with a value of ~0.43. This confirms that sample annealing is sufficient to diffuse inherent defects in h-BN flakes without electron beam irradiation. However, single photon purity from the $E_1$ point is less, as inferred from the shallow $g^2(\tau)$ dip. In addition, the emission from point $E_1$ is not stable for a long time and shows photo-blinking, as shown in Figure 4d (upper panel). We observe significant fluctuations in photon counts, which shows that emitter is not photo-stable, as seen in Figure 4d (lower panel), which is undesirable for any quantum application. Furthermore, the $E_1$ point is permanently photo-bleached once illuminated for a long time or excited with higher excitation power. Thus, although it is possible to obtain an isolated, single photon emitter through high-temperature annealing in a pristine h-BN flake, the resulting single photon emission is not stable with shallow $g^{(2)}(\tau)$ dip at $\tau = 0$. Hence, a low-energy electron beam irradiation and subsequent high-temperature annealing are required to achieve photo-stable, reproducible, high-purity on-demand single photon sources with a high emission rate. However, we do not observe quantum emitters from pristine samples (the non-irradiated and not-annealed sample) i.e. *F3* sample as elaborated in the supplementary information S8. The Raman peak at 573 nm due to h-BN is observed in the emission spectra from pristine h-BN sample.

## III. CONCLUSIONS

To summarize, we have studied quantum light emission from an isolated 'single' stable single-photon emitter in h-BN flakes at room temperature. The low-energy electron beam irradiation and subsequent high-temperature annealing are crucial to creating site-specific, photo-stable, single-photon emitters. The quantum emitter shows a prominent ZPL emission peak compared to PSB emission with a high Debye-Waller factor of ~50±10%. Accordingly, the Huang-Rhys factor shows a significantly lower value (~0.6 << 1) at room temperature, implying minimal electron-phonon interaction in emission process. The second-order auto-correlation measurement confirms photon anti-bunching with normalized correlation count ~0.25 at delay time $\tau = 0$ with a lifetime of 2.46±0.1 ns. Such fast single photons with minimal decoherence in the system are essential for quantum communications and efficient spin-photon interface. The presence of a stable single-photon emitter in our system is attributed to the creation of a carbon trimer defect complexes in h-BN samples. The extensive quantum optical studies on 'only annealed' flakes suggest the importance of low-energy electron beam irradiation to create stable single photon emitters in h-BN flakes. Although we find substantial evidence of single photon quantum emitters in 'only annealed' flakes, these emitters experience photo-blinking effects and subsequently get photo-bleached with time, indicating them as an inefficient source of single photons.

**Acknowledgments and Funding:**

We thank Rahul Dhankhar and Debdip Guchait for assisting in the experiments. PJ acknowledges the institute fellowship from IIT Jammu. BC acknowledges the central instrumentation facility at IIT Jammu. RVN thanks the central research facilities at IIT Ropar. The authors thank the financial support from Department of Science and Technology (DST), India [DST-ICPS [DST/ICPS/QuST/Theme-2/2019/General], DST-SERB [SB/SJF/2020-21/05], Swarnajayanti Fellowship (DST/SJF/PSA-01/2019-20), and DST-SERB[SRG/2020/000563]. AB acknowledge the institute postdoctoral fellowship from IIT Ropar.

**Data Availability Statement:**

The data that support the findings in this study are available from the corresponding author upon reasonable request.

**Supporting Information:** Details of sample fabrication; quantum optical experiments; emission spectra from many samples; Raman spectra measurements; additional polarization-dependent emission measurements; additional emitters from sample *F1* type, spatial-dependent emission measurements, and studies on pristine h-BN flakes.

**Notes:** The authors declare no competing financial interest.

# Site-specific stable deterministic single photon emitters with low Huang-Rhys value in layered hexagonal boron nitride at room temperature

Amit Bhunia[1*], Pragya Joshi[2*], Nitesh Singh[1], Biswanath Chakraborty[2]
and Rajesh V Nair[1]

[1]Laboratory for Nano-scale Optics and Meta-materials (LaNOM), Department of Physics, Indian Institute of Technology Ropar 140001, India

[2]Department of Physics, Indian Institute of Technology Jammu, Jagti, NH-44, Jammu and Kashmir 181221, India

*contributed equally

### S1. Sample fabrication and creation of emitters

Few layers of h-BN flakes were mechanically exfoliated from a h-BN crystal (M/s. 2D Semiconductors, USA) using the standard scotch tape method onto a $SiO_2$/Si substrate with 285 nm of the oxide layer. The sample is examined in an optical microscope to locate the growth and position of the desired h-BN flake. To create defects in the h-BN flake, the sample is irradiated with an electron beam in a field emission scanning electron microscope with an acceleration voltage of 20 keV and a beam current ~ 6 µA. The beam is focused on the targeted h-BN flake at high magnification (~5,00,000X) for 20 seconds. Sample was subsequently annealed in argon atmosphere at $850^0$ C for 2 hours under continuous argon flow. The flow rate was adjusted to be at 1 bubble per second. We have observed that increasing the irradiation time beyond 20 seconds causes h-BN to show graphitic behaviour with a significant enhancement in the 1580 cm$^{-1}$ peak. Annealing for less than 2 hours did not produce stable emitters and such emitters die out under illumination. There always exists a finite probability of organic contaminations on the surface of the sample during fabrication process. Thanks to the high temperature annealing methods which manifest to eliminate any such contamination (e.g. hydrocarbon, undesired residue of chemicals, etc.) from the surface happened during the fabrication time. Here, we consider three different types of h-BN flakes as mentioned also in the main text: (i) 'Irradiated and annealed' flakes (*F1 sample*), (ii) 'only annealed' flakes (*F2 sample*), and (iii) 'pristine or not annealed' flakes (*F3 sample*).

## S2. Experimental details

### S2A. Scanning confocal imaging and emission measurements.

To probe the photonic properties of isolated quantum emitters on h-BN flake at room temperature, we have developed an indigenous scanning confocal microscopy and spectroscopy set-up as seen in Figure S1. We use a 532 nm picosecond diode laser (LDHPFA 530 L, PicoQuant) as an excitation source having a pulse width of 80 ps and repetition rate of 10 MHz (unless mentioned otherwise) and the same laser is also used in CW mode. The laser beam size from the laser head is increased using beam broadening optics and subsequently, the light is directed toward the sample using mirrors and a 90:10 polarizing beam splitter (PBS). The excitation source illuminates completely the microscope objective of 60X magnification and numerical aperture of 0.95 (CFI Plan Apo Lambda 60XC) and subsequently, the excitation beam is focussed on the sample. The spot size of the excitation laser beam on the sample is approximately ~1.0 µm in diameter. The sample is kept on an *x-y-z* actuator with a spatial resolution of 50 nm (M/S, Physik Instrumente, GmBH). The excitation power at the entrance of the objective is varied depending upon measurement requirements using neutral density filters. The typical power used to excite the sample using CW laser is ~0.5 *mW* (unless mentioned otherwise) measured using an Ophir power sensor just before the microscope objective.

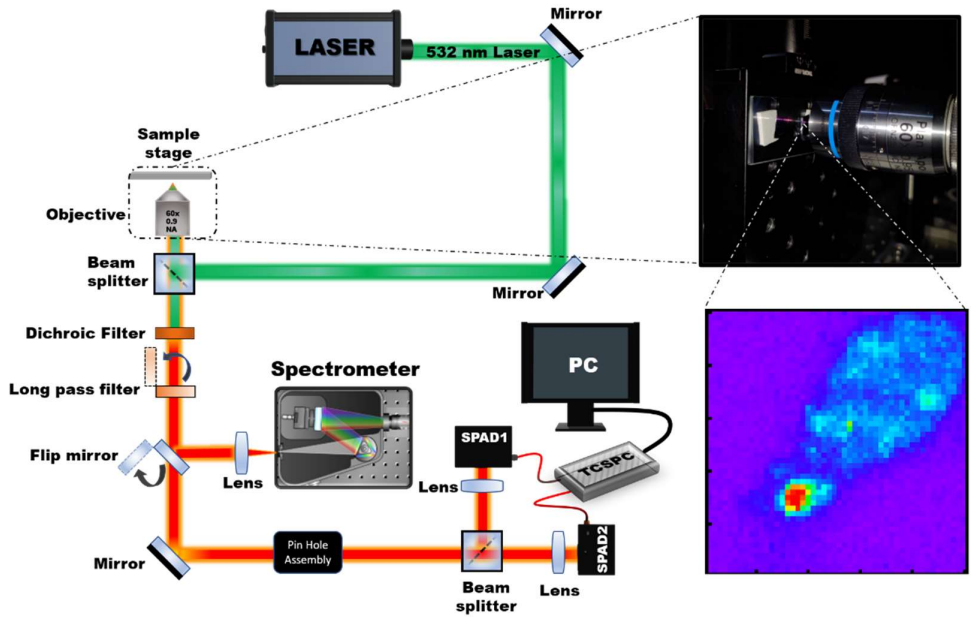

**Figure S1.** The schematic of the home-built confocal microscopy setup and quantum optical experimental setup which is used to measure the emission spectra, decay rate, and second-order intensity-intensity correlation.

The emitted light from the sample is collected using the same objective and passed through a 532 nm dichroic mirror and 550 nm long pass filter to separate out the excitation beam. The emitted light is then spatially filtered using a 50 μm pinhole assembly and detected using a fiber-coupled single-photon avalanche photodiode (Excelitas, excelitas technology). The decay rate profile of the quantum emitter is studied using a time-correlated single photon module (HydraHarp 400, PicoQuant) synchronized with an avalanche photodiode (APD) and the excitation laser control unit. For the spectral measurement, the emitted light is directed towards a spectrometer with the help of a flip mirror placed in the emission path. We used the spectrometer (M/S. Andor Technology Ltd.) with a blazed grating of maximum operational efficiency at 550 nm having 900 grooves/mm. The spectrometer is equipped with an electron-multiplying charge-coupled device (EMCCD) for spectral characterization. The EMCCD resolution for the particular measurement is ~0.14 nm at 580 nm for a fixed slit width of 50 μm.

**S2B. $g^{(2)}(\tau)$, polarization-dependent emission, and Raman spectra measurements.**

Second-order auto-correlation $g^{(2)}(\tau)$, measurements are performed using a home-built Hanbury Brown and Twiss set-up as seen in Figure S1. Here, additionally, the emitted light is split into two paths using a 50:50 plate beam splitter (Thorlabs) and subsequently detected using two fiber-coupled APDs (Excelitas, excelitas technology) in the exact perpendicular configuration. The $g^{(2)}(\tau)$ measurements are performed using both pulsed as well as CW lasers to confirm the repeatability and reproducibility of single photon emission. The measured correlation counts are analyzed with the help of dedicated Hydraharp 400 time tagging software. The polarization-dependent measurements are carried out by inserting Glan-Thompson (GT) linear polarizers in the excitation and emission paths. The angle-dependent emission spectra are recorded by rotating the analyzer angle. The micro-Raman measurements were carried out using a commercial Raman spectrometer (LAB RAM HR800) from Horiba. All the optical characterization is done at room temperature at ambient air conditions.

**S3. Raman spectra of the *F1* sample**

Raman spectroscopy is an important tool to characterize the implicit and unique vibrational modes associated with materials. The Raman spectra measurement are done using 532 nm laser excitation to understand the distinct Raman peaks associated with few-layer h-BN flake (*F1* sample). The measured

Raman spectra show a Raman peak shift at 1365 cm$^{-1}$ that is assigned to the E$_{2g}$ vibrational mode of h-BN, as shown in Figure S2.

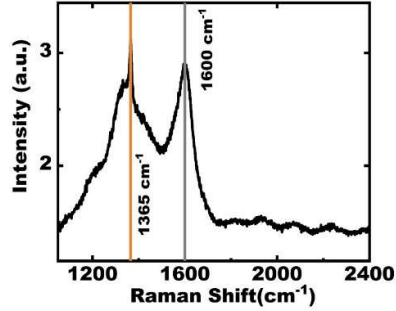

**Figure S2.** Raman shift measurement shows the typical Raman peak appearing at ~1365 cm$^{-1}$ of h-BN. As such we do see this Raman peak shift in most of the h-BN flakes discussed in the manuscript.

The peak at 1365 cm$^{-1}$ rather confirms that this is a few layers of the h-BN sample. In addition to the well-known 1365 cm$^{-1}$ peak of h-BN, we observe another prominent Raman peak at 1600 cm$^{-1}$ which resembles like the G-band Raman peak of graphite or graphene[1]. Moreover, it is understood that carbon is ubiquitously present in layered h-BN, usually incorporated unintentionally during the growth or post-processing of the h-BN materials[2]. Therefore, we can infer that due to the presence of carbon impurity in the h-BN sample, there might occur some partial graphitization which effectively contributes to the G-band Raman peak in the spectra[3].

**S4. Polarization-dependent emission measurements from the *F1* sample**

Controlling the polarization state of the emitted light from an isolated quantum emitter plays an essential role in quantum information processing.[4] Moreover, the polarization of emitted photons from a quantum emitter can be utilized and manipulated to build a polarization entanglement state, which paves the way for secure data communication or quantum cryptography.[5] Hence, it is a prerequisite to study the dipolar orientation of the quantum emitter inside the solid-state system, which reveals the dipolar nature of the quantum emitter. Figure S3a represents the analyzer angle-dependent integrated emission intensity at the PSB emission peak. The polarization of the excited laser is kept at a fixed angle and the optic axis of the linear polarizer in the excitation path is aligned vertically to the experimental frame of reference. Usually, the ZPL emission peak mainly contributes to dipolar emission whereas the PSB peak hardly shows any such behavior. However, we observe prominent dipolar emission for PSB emission peaks, and the angular intensity distribution exhibit a more pronounced dipolar nature in the emission process, in comparison to the ZPL peak (Figure 2d in the main manuscript). The average polarization angle $\langle\theta\rangle$ for the PSB peak is estimated to be ~26±0.7⁰. The value apparently represents the orientation of a single quantum emitter with respect to the analyzer axis. Also, the estimated visibility appears around 77% for PSB-related-emission. Here, we observed a significantly higher degree of polarization in PSB-related emission peaks in comparison to ZPL emission peaks. Therefore, the observed dipolar behavior from PSB emission certainly poses an important question regarding the underlying emission mechanism and involved electronic levels contributing to PSB emission.

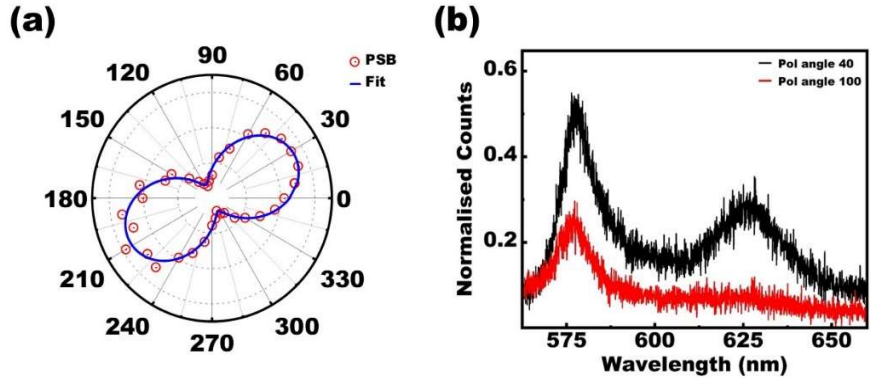

**Figure S3.** (a) The polar plot represents integrated emission intensity at the PSB peak as a function of the analyzer angle while the excitation laser polarization is kept fixed at the optic axis of the polarizer. The measured emission intensity (symbols) is fitted with $cos^2\theta$ function (solid line) which gives the polarization angle of the emitter as 26±0.7⁰. The radiation pattern is more resemble like a dipole pattern in comparison to ZPL related angular emission pattern shown in Figure 2d. The degree of polarisation is estimated to be ~ 77% for PSB-related emission (b) The contrasting emission behavior of quantum emitter depends upon the angle of analyser. Based on the angle of the analyser, it is possible to suppress the PSB-related peak where the emission due to ZPL still remains prominent.

Figure S3b shows the measured emission spectra from the $Q_1$ point on the *F1* sample for two different analyzer angles. It is evident that the PSB-related emission peak is almost vanished for a specific analyzer angle whereas the ZPL-related emission peak is intact at ~ 578 nm. The spectral features certainly point towards an interesting observation, however, the investigation might be carried forward to reveal the underlying mechanism of reduced emission at PSB. Here, we would like to mention that a polarizing beam splitter of 90:10 ratios is placed in the excitation path of the emission setup to control the excitation polarization. It might happen that the beam splitter has played a typical role to modify light polarization in the emission path. Therefore, we foresee that PSB-suppressed pure ZPL emission could be possible to achieve in few-layer h-BN sample by properly tuning the angle of the polarizer, paving the way for better quantum technological applications like spin-photon interface and coherent spin manipulation.

**S5. More quantum emitters from different flakes belongs to *F1* sample category.**

We have fabricated multiple h-BN flakes similar to the *F1* type samples. This particular batch of samples are fabricated at different time in comparison to *F1* sample described in the main text. The samples are rigorously studied to substantiate deterministic, stable single photon emitters of low HR factors. Figure S4 represents the characteristic emission properties of another quantum emitter (point 1) measured on another flake *(F1$_a$ sample)*. Here the single emitter is located at the sample center (marked using black square box) as we intentionally irradiated at the center using the electron beam to create defects on the sample. This *F1$_a$* sample also shows the similar emission properties like the *F1* sample as discussed in the main manuscript. Figure S4a shows the typical confocal map of *F1$_a$* sample which shows the isolated quantum emitter (marked using square box) measured using few hundred μW of 532 nm laser. The confocal image confirms only a single isolated and distinct emitter on the flake instead of cluster of multiple emitters, similar to the *F1* sample. Emission spectra measured from this point shows distinct ZPL signature at 576 nm with small hump due to PSB emission at ~626 mm as shown in Figure S4b.

The measured anti-bunching curve ($g^2(\tau)$) shows the single photon emission with characteristics anti-bunching dip around~ 0.5 at $\tau=0$ as seen in Figure S4c. The decay profile of the emitted photon is measured and fitted using an exponential function to estimate the lifetime as shown in Figure S4d. The estimated lifetime of the emitter is 2.47 ±0.3 ns, in agreement with lifetime value of *F1* sample shown in main manuscript (Figure 3b). The photo-stability studies of the quantum emitter in the *F1a* sample shows stable photon counts for more than 5 minutes as illustrated in Figure S4e. Our result emphasizes the creation of stable quantum emitter using site-specific electron beam irradiation and subsequent annealing of sample belonging to the *F1* sample category in comparison to the non-irradiated samples as discussed in the main text.

To further corroborate on the dipolar nature of the emitter similar to the results shown in the main manuscript, we have carried out the extensive polarisation-dependent emission measurement at the ZPL wavelength. Figure S4f shows the dipolar behaviour of the emitter having a polarisation angle of 28±0.8⁰. Therefore, these results from another quantum emitter on *F1a* sample belongs to the *F1* sample category confirms that it is possible to recreate and reproduce the quantum emitter controllably and deterministically as per the demand.

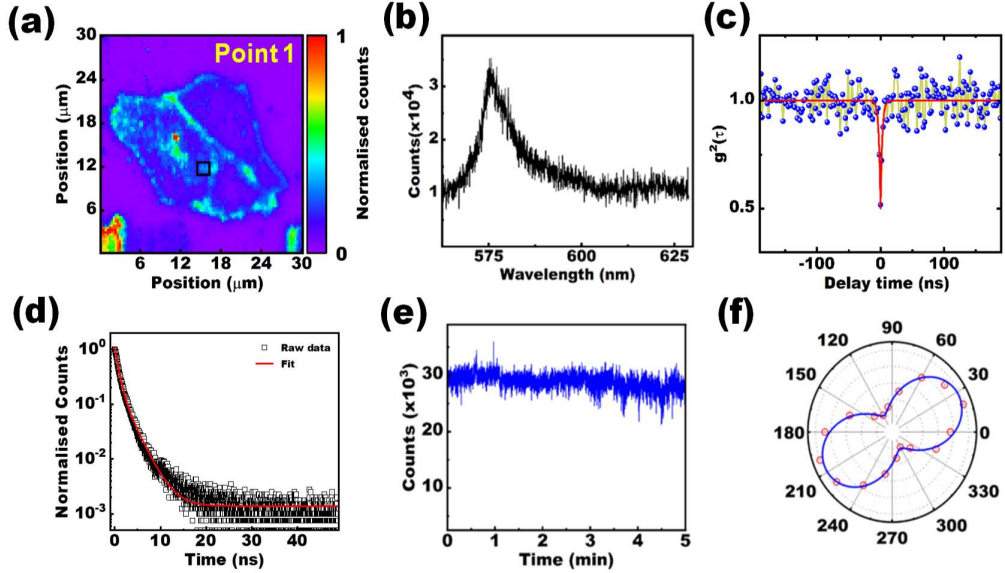

**Figure S4.** (a) Confocal map of the *F1a* sample measured with 532 nm excitation. The black square encircles the quantum emitter (point 1) on the flake. (b) The measured emission spectra from the encircled region shows prominent ZPL transition at 576 nm. (c) The corresponding anti-correlation measurement shows the photon anti-bunching dip at zero-time delay, suggesting the single photon emission. The red solid line is the single exponential fit to the measured data, similar to the analysis given in the main manuscript. (d) The measured emission decay profile is fitted with exponential decay function (solid red line) provides emission lifetime of 2.47 ± 0.3 ns. (e) Photo-stability of the emitter is shown, which confirms the stable emission from the isolated emitter. (f) Polarisation-dependent optical emission properties of the quantum emitter, which shows the dipolar nature of the emission pattern owing to the single emitter.

In Figure S5, we demonstrate the emission properties and single photon emission from another three samples like the *F1* sample. All the sample show similar emission spectra with ZPL line at 576±2 nm accompanied with PSB emission around 620 nm. The single photon emission properties with the characteristics dip in the measured anti-bunching curve at τ=0 delay for all the samples as seen in Figure S5. Finally, we summarise the relevant optical parameters in Table 1 for all the five single photon emitters studied in different samples, which belongs to *F1* sample category. It is evident from the Table 1 that the ZPL peak wavelength varies ~ ±2 nm dependent upon different locations and experimental conditions. The ZPL wavelength is very much intact at ~576 nm, which indicates the sample stability with similar kinds of defect created in all the samples. The estimated HR value for all the samples is much less than unity, indicating minimal electron-phonon interaction during the emission process which results in larger fraction of emission intensity funnelled into the ZPL transition.

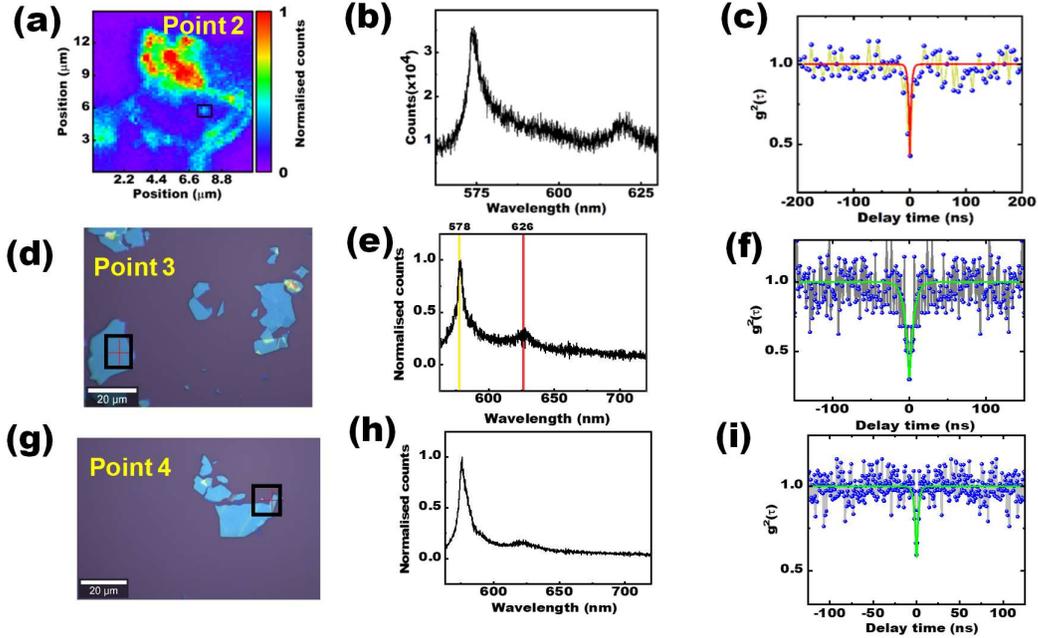

**Figure S5.** Left panel: (a), (d) and (g) represent the location of the quantum emitters at different region on each samples (marked with square box) by focusing the electron beam at different location on each samples. Point 2 is located in the confocal map (a), point 3 is located in the microscope image (d) and point 4 is located in the microscope image (g). Middle Panel: (b), (e) and (h) show the measured emission spectra using 532 nm laser excitation from the emitters corresponding to points 2, 3 and 4, respectively. Right panel: (c), (f), and (i) show the measured $g^2(\tau)$ curve with characteristics single photon emission dip at τ=0 time delay from points 2, 3, and 4 respectively.

**Table S1: Emission parameters corresponding to different quantum emitters.**

| Quantum emitters | ZPL Wavelength | Measured HR factor | $g^2(\tau=0)$ |
|---|---|---|---|
| Point 1 | 576 nm | 0.68 | 0.52 |
| Point 2 | 575 nm | 0.76 | 0.42 |
| Point 3 | 578 nm | 0.61 | 0.3 |
| Point 4 | 576 nm | 0.48 | 0.59 |
| $Q_1$ point (shown in the main manuscript) | 578 nm | 0.6 | 0.25 |

## S6. Spatial-dependent emission properties of the *F1* sample

Figure S6a shows the spatial-dependent emission spectra measured on the *F1* sample. The inset of Figure S6a shows the confocal image of the *F1* sample with two distinct points marked $Q_1$ and $d_1$ on the sample. It is evident that the $Q_1$ point shows emission spectra with ZPL at 578 nm and PSB-related emission at 626 nm. The Raman peak of h-BN is feeble (a narrow peak ~ 573 nm (1365 cm$^{-1}$)) in the measured emission spectra from the $Q_1$ point due to substantial spontaneous emission. However, point $d_1$ on the *F1* sample shows very weak emission but with well-resolved prominent Raman peaks at 573 nm and 581 nm, which suggest that the $d_1$ region contains more carbon-related defect clusters and hence minimal emission features from h-BN.

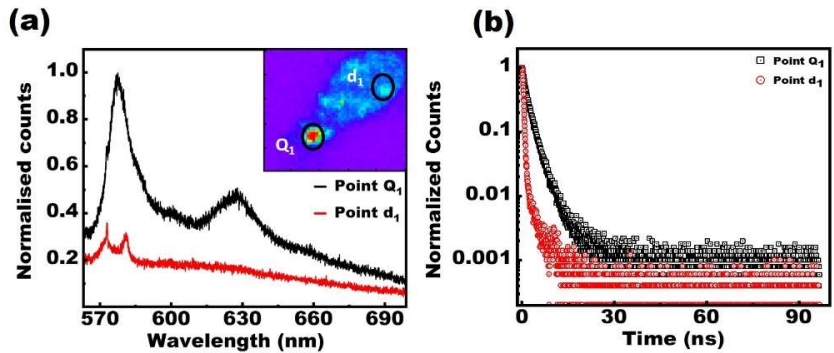

**Figure S6.** (a) The comparison between measured emission spectra from two distinct locations on the *F1* sample (irradiated and annealed) are compared. The inset shows the confocal image of the *F1* sample with two distinct marks ($Q_1$ and $d_1$ points). An intense emission spectrum that features the ZPL and PSB emission peaks from the $Q_1$ point on the sample is seen. However, the spectra obtained from any other arbitrary location on the flake ($d_1$ point) show only Raman-related scattering peaks. (b) The excited state emission decay curves from the $Q_1$ and $d_1$ points show distinct decay profiles. The point $d_1$ initially decays quite fast indicating the scattering-related transition involved in the process. The $Q_1$ point shows a comparatively slower decay rate and an estimated lifetime of 2.54 ns implying electronic transition is involved in the emission process.

Besides, this study indicates the importance of spatial mapping of emission from samples to identify regions wherein the quantum emitter is located. Moreover, spectra measured from the $Q_1$ point are substantially distinct in comparison to typical Raman-related scattering emission measured from the $d_1$ point. Figure S6b shows the measured excited state decay profile of the quantum emitter located at $Q_1$ point which is compared with the decay profile measured from the $d_1$ point of the sample. The emission decay corresponds to $d_1$ point is a fast process with an excited state lifetime of ~ 0.4 ns implying the possible link to a scattering-related emission process. However, we observe significantly different decay behavior from the emitter located at the $Q_1$ point with an excited state lifetime of 2.54 ns. Thus, the decay rate measurements from different locations on the h-BN flake are insightful to identify the presence of a quantum emitter in the sample.

**S7. AFM image of *F2* sample**

The thickness of the *F2* sample is measured using the atomic force microscopy (AFM) technique. Figure S7 depicts the AFM image of the *F2* sample along with the height profile corresponding to a particular location (black dashed line). The estimated height of the sample is measured to be ~25±1 nm. The average height of the *F2* sample is matching with average height of *F1* sample as discussed in the main text.

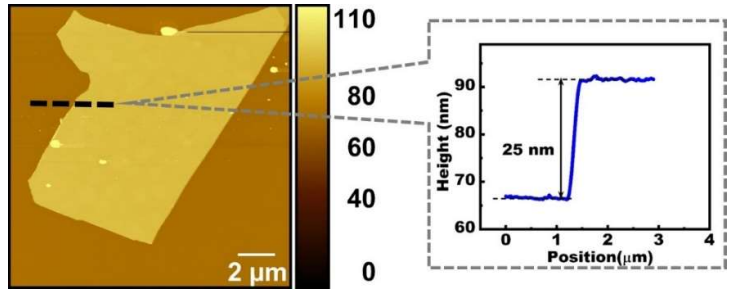

**Figure S7.** The AFM image of the *F2* sample along with the height profile from a particular location (black dashed line). The flake thickness is nearly constant throughout the sample which is ~25±1 nm. For better visualization, the arrow and the dotted lines are drawn at the maximum and minimum.

**S8. Comparison between annealed (*F2* sample) and not annealed (*F3* sample) h-BN flakes**

Figure S8 shows the emission spectra measured for annealed sample (*F2* sample) and the not annealed sample (*F3* sample) having equivalent optical contrast (i.e. thickness) under optical microscope. Both *F2* and *F3* samples are not irradiated. We observe a significant difference in the measured emission spectra from *F2* and *F3* samples. Two sharp and intense spectral peaks are observed at ~573 nm and 581 nm for *F2* sample whereas such prominent peaks are absent from *F3* sample except for a small transition peak around 573 nm. The measured intense dual peaks are detected from most of the regions of the *F2* sample except from the $E_1$ point discussed in the main manuscript. Furthermore, we find stronger evidence to compare those two sharp peaks with Raman-related transition. The Raman shifted peaks once converted, appear at ~ 1363 cm$^{-1}$ and 1580 cm$^{-1}$ corresponding to ~573±1 nm and ~580±1 nm wavelength, respectively. These two Raman peaks are assigned to $E_{2g}$ mode of layered h-BN and G-band of graphite/graphene due to the presence of carbon-related defects, respectively. However, we see an intensity-suppressed Raman peak of h-BN for *F3* sample and do not see any carbon-related Raman peak. Therefore, our emission measurements summarize the importance of annealing to activate electronic as well as vibronic transition channels which effectively contribute to an intense emission peak along with an extra carbon-related Raman peak in the emission spectra.

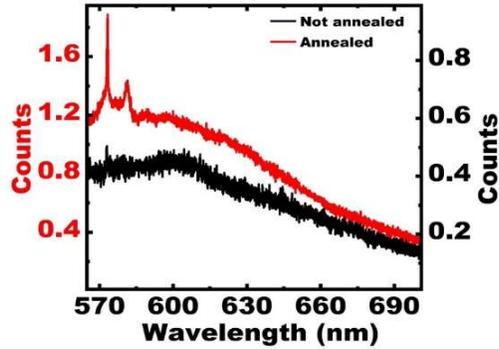

**Figure S8.** The measured emission spectra from both annealed (*F2*) and not annealed (*F3*) h-BN flakes with contrasting spectral features. The *F2* sample show two sharp prominent peaks at ~573 nm and ~580 nm. However, *F3* sample does not show any significant emission features except a small peak observed at ~573 nm.

## S9. Fitting methods and estimated parameters

Here, we discuss the different fitting mechanisms and models to extract the relevant parameters from the experimentally obtained results. It always remains a challenge to perfectly fit an experimental data with a theoretical model and it has been realized as a rigorous process to completely match the measured data. In Figure S9, we show the fitting of the emission spectra as discussed in the main manuscript (Figure 2b). Here, we use the multiple peak fitting function available in Origin graph plotting software. Both the ZPL and PSB are fitted with Lorentz peak function. We varied and optimized the corresponding fitting parameters to match the best overlapping of the experimental data and Lorentz peak functions. From the fitted curves, we extract the relevant parameters like area under the curve and line width, which are used to estimate the DW values.

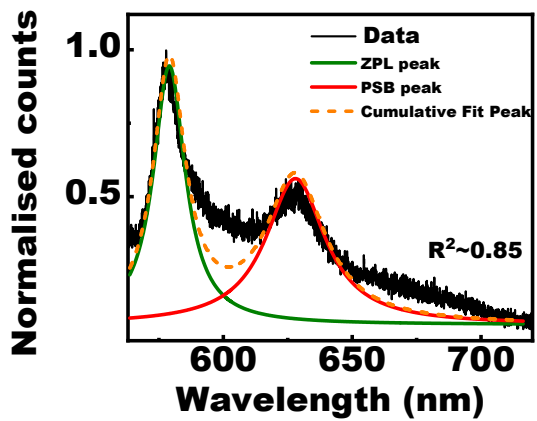

**Figure S9.** The emission spectra is fitted to estimate the relevant spectral information. Lorentz peak function is used to fit the both ZPL and PSB peaks. Corresponding R-square value appears to be~ 85%.

In Figure S10, we discuss the details of the fitting to the measured photon anti-bunching curve as shown in Figure 3a of the main manuscript. Here we use single exponential function to fit the experimental data. The experimentally obtained raw data is normalised to unity and background corrected for better visibility and understanding. For background correction,[6] we use the following relations $g^2(\tau') = \frac{(g^2(\tau) - (1-\rho^2))}{\rho^2}$, where $g^2(\tau')$ is the background corrected auto-correlation function. $\rho^2 = \frac{S}{S+B}$ where $S$ represents the emission counts from the emitter, and $B$ is the background emission counts. Inset of the Figure S10 shows the fitting parameters and the estimated values of the lifetime.

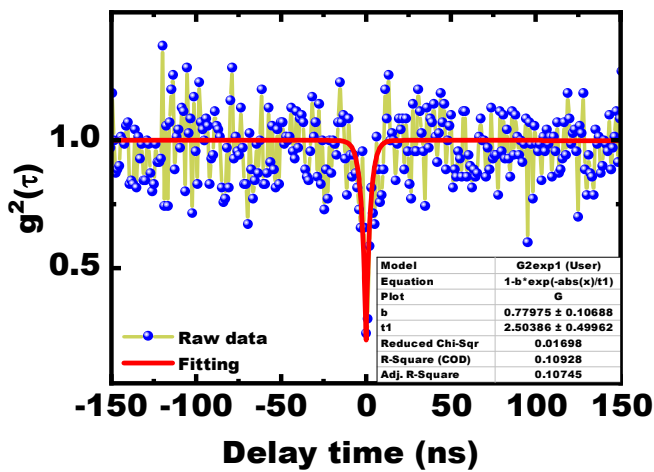

**Figure S10.** The auto-correlated photon counts are fitted with a single exponential function. The corresponding fitting parameters and equation are shown in the inset of the figure.